\documentclass{jfm}
\usepackage{graphicx}
\usepackage{epstopdf, epsfig}

\usepackage{amsmath,amssymb, bm}
\newcommand{\hG}{\hat{G}}
\newcommand{\hPhi}{\hat{\Psi}}

\newcommand{\bOmega}{\bm{\Omega}}

\newcommand{\bD}{\bm{D}}
\newcommand{\be}{\bm{e}}
\newcommand{\bE}{\bm{E}}

\newcommand{\bx}{\bm{x}}

\newcommand{\bu}{\bm{u}}

\newcommand{\bq}{\bm{q}}

\newcommand{\bomega}{\bm{\omega}} 
 
\newcommand{\odiff}{\text{d}}

\newcommand{\dbx}{\odiff\bx}
\newcommand{\dbq}{\odiff\bq}
\newcommand{\avg}[1]{\langle #1 \rangle}
\newcommand{\bavg}[1]{\big\langle #1 \big\rangle}
\newcommand{\Bavg}[1]{\Big\langle #1 \Big\rangle}

\newcommand{\bj}{\bm{j}}
\usepackage{xcolor}

\shorttitle{Rotational Taylor dispersion in linear flows}
\shortauthor{Z. Peng}

\title{Rotational Taylor dispersion in linear flows}
\author{Zhiwei Peng
\corresp{
\email{zhiwei.peng@ualberta.ca}
}}
\affiliation{Department of Chemical and Materials Engineering, University of Alberta, Edmonton, Alberta T6G 1H9, Canada}

\begin{document}

\maketitle

\begin{abstract}
The coupling between advection and diffusion in position space can often lead to enhanced mass transport compared with diffusion without flow. An important framework used to characterize the long-time diffusive transport in position space is the generalized Taylor dispersion theory. In contrast, the dynamics and transport in orientation space remains less developed. In this work we develop a rotational Taylor dispersion theory that characterizes the long-time orientational transport of a spheroidal particle in linear flows that is constrained to rotate in the velocity-gradient plane. Similar to Taylor dispersion in position space, the orientational distribution of axisymmetric particles in linear flows at long times satisfies an effective advection-diffusion equation in orientation space. Using this framework, we then calculate the long-time average angular velocity and dispersion coefficient for both simple shear and extensional flows. Analytic expressions for the transport coefficients are derived in several asymptotic limits including nearly spherical particles, weak flow and strong flow.  Our analysis shows that at long times the effective rotational dispersion is enhanced in simple shear and suppressed in extensional flow. The asymptotic solutions agree with full numerical solutions of the derived macrotransport equations and results from Brownian dynamics simulations. Our results show that the interplay between flow-induced rotations and Brownian diffusion can fundamentally change the long-time transport dynamics. 
\end{abstract}

\begin{keywords}
colloids, dispersion, microscale transport
\end{keywords}

\section{Introduction}
Transport and mixing of solutes or particles in the presence of hydrodynamic flows are important for various biological and industrial processes. For micron-sized particles immersed in flows, the coupling between advection and diffusion can often lead to enhanced mass transport as compared with diffusion without flow. A classical example of such a coupling effect is the Taylor dispersion of Brownian solutes in pressure-driven channel flows~\citep{Taylor1953,Taylor1954b,Taylor1954a, Aris1956}. Brownian motion allows the solute particles to migrate across streamlines and then  be advected downstream with different velocities. At long times, the coupling between transverse diffusion and longitudinal advection gives rise to diffusive transport of the solutes with an effective longitudinal dispersion coefficient that can be much larger than the bare diffusivity of the solute particle. Since the work of \citet{Taylor1953}, a generalized Taylor dispersion (GTD) framework \citep{frankel1989foundations} has been developed to accommodate a wide range of transport problems including complex geometries, chemical reactions, spatial and/or time periodicity and active particles~\citep{brenner1980dispersion,SHAPIRO1986,shapiro87,shapiro1990taylor,morris_brady_1996,hill2002taylor,bearon2003extension,manela2003generalized,zia_brady_2010,takatori2014swim,Burkholder17,alonso2019transport,jiang_chen_2019,PB2020}. More recently, longitudinal dispersion of elongated Brownian rods in  a two-dimensional Poiseuille flow has been considered~\citep{kumar2021taylor,khair2022taylor}.

In contrast to the extensive study of the long-time effective transport of particles in position (linear) space, the dynamics and transport of particles in orientation space remains relatively less developed. For spherical or `point' particles, the consideration of the orientational dynamics is often unnecessary. For anisotropic particles, their orientational dynamics plays a role in the overall dynamics and rheology of the suspension composed of the particles and the fluid~\citep{leal1971effect,hinch1972effect,khair2016suspension}. A typical example is the orientational dynamics of an isolated spheroid in simple shear. Under shear, the orientation of the spheroid undergoes complex dynamics governed by Jeffery's equation~\citep{Jeffery1922}. As a result, a Brownian spheroid in simple shear experiences both rotational diffusion and angular advection that is non-uniform. An interesting question we wish to consider is: Does the coupling of advection and diffusion in orientation space lead to enhanced rotational transport? 

Using experiments and particle-based simulations, \citet{Leahy2013} showed that advection-diffusion coupling indeed results in enhanced rotational diffusion at long times for an axisymmetric particle under shear.  In a later paper~\citep{leahy_koch_cohen_2015}, a continuum theory is developed to calculate the time-dependent orientation distribution for non-spherical axisymmetric particles confined to rotate in the velocity-gradient plane, in the limit of weak diffusion or large P\'eclet number (see sec.~\ref{subsec:simple-shear} for the definition). In this limit, a coordinate transformation is discovered and used to map the orientation dynamics to a diffusion equation, which ultimately allowed the calculation of the long-time rotational dispersion coefficient. Furthermore, a remarkably simple analytic expression is obtained for the dispersion coefficient in the large-P\'eclet-number limit. In comparison to the classical Taylor dispersion, \citet{leahy_koch_cohen_2015} concluded that their theoretical consideration does not fit nicely under the canonical GTD framework. 

In this work we show that the flow-enhanced rotational transport in the velocity-gradient plane can be treated as a GTD in orientation space. To setup the system for such a treatment, the key step is to consider the unbounded angular displacement $\varphi$ rather than the orientation angle $\phi$, which is bounded to an interval of length $2\pi$. With this, one can then break down the unbounded displacement into an infinite sequence of cells, each of which has a length of $2\pi$. In the language of GTD, one then identifies the cell index $j \in \mathbb{Z}$ as the global coordinate and $\phi$ as the local coordinate. The derived GTD theory works for generic linear flows and for arbitrary P\'eclet numbers. In the large-P\'eclet-number limit, we show that our asymptotic result agrees with that obtained by \citet{leahy_koch_cohen_2015} for steady simple shear. Our results from the GTD theory is validated against Brownian dynamics (BD) simulations. 

In $\S$ \ref{sec:formulation}, starting from the Smoluchowski equation governing the orientational dynamics of a spheroidal particle, we develop the GTD formulation for generic linear flows. Following \citet{leahy_koch_cohen_2015}, the particle is constrained to rotate in the velocity-gradient plane. In $\S$ \ref{sec:results}, we consider the long-time rotational transport in simple shear and extensional flows. The transport coefficients are calculated using  perturbation expansions in both small- and large-P\'eclet-number limits. The results obtained in these asymptotic limits are compared with numerical solutions of the macrotransport equations and results from BD simulations. Lastly, we conclude the paper in $\S$ \ref{sec:conclusion}.

\section{Problem formulation}
\label{sec:formulation}
\subsection{The Smoluchowski equation}
Consider a spheroidal particle immersed in a linear ambient flow field in an unbounded, incompressible Newtonian fluid. The particle is sufficiently small so that inertia effects are neglected and the fluid is in the Stokes regime. The particle is subject to rotational Brownian motion and no external torque is applied. In the absence of Brownian motion, the time evolution of the unit orientation vector $\bq$ ($|\bq|=1$) of the particle is governed by Jeffery's equation \citep{Jeffery1922}:
\begin{equation}
    \label{eq:Jeffery}
    \frac{\dbq}{\odiff t} = \bOmega\times \bq \quad \text{and}\quad \bOmega = \frac{1}{2}\bomega + B \bq \times \left( \bE \bcdot \bq \right).
\end{equation}
Here $\bOmega$ is the instantaneous angular velocity, $\bomega = \bnabla \times \bu$ is the vorticity vector, $\bE = \frac{1}{2}\left( \bnabla \bu  + ( \bnabla \bu)^\intercal \right)$ is the rate-of-strain tensor, $\bu$ is the ambient flow field and $B = (r^2-1)/(r^2+1) \in [0,1)$  is the Bretherton constant that characterizes the non-sphericity \citep{bretherton1962motion}, with $r$ the aspect ratio of the spheroid. For a sphere, $r=1$ and $B=0$. For an infinitely thin rod, $r \to \infty$ and $B \to 1$. 

With Brownian motion, a statistical mechanical description is required. To this end, we define the orientational probability density function $\Psi(\bq, t)$, which satisfies the total conservation condition $\int_{\mathbb{S}} \Psi(\bq, t) \dbq = 1$ at (any) time $t$.  Here, $\mathbb{S} = \{\bq\, \lvert\, \bq\bcdot\bq=1 \}$ denotes the unit sphere of orientations. The orientational probability density function is governed by the Smoluchowski equation \citep{Brenner1974,doi1988theory}
\begin{equation}
\label{eq:smol-gen}
    \frac{\partial \Psi }{\partial t} + \bnabla_R\bcdot \bj_R =0,
\end{equation}
where $\bj_R = \bOmega \Psi - D_R \bnabla_R \Psi$ is the rotational flux vector, $\bnabla_R = \bq \times \partial/\partial \bq $ is the rotational gradient operator and $D_R$ is the rotational diffusivity. 

We note that \eqref{eq:smol-gen} can be treated as  a marginalization of the full probability density function $P(\bx, \bq, t)$ that describes the joint distribution of the particle in both position and orientation space, where $\bx$ is the position vector of the particle centre. This full probability is governed by 
\begin{equation}
    \frac{\partial P }{\partial t} + \bnabla\bcdot\bj_T + \bnabla_R\bcdot \bj^\prime_R =0,
\end{equation}
where, for a  Brownian particle $\bj_T =\bu P - \bD_T(\bq) \bcdot\bnabla P$, and $\bj^\prime_R = \bOmega P - D_R \bnabla P$. Here, $\bD_T$ is the translational diffusivity of the particle, which is a function of $\bq$ for a spheroid. It is clear that $\Psi(\bq,t) = \int P(\bx, \bq, t)\dbx$. For active (self-propelled) Brownian particles with a constant swim speed $U_s$, an additional term $U_s\bq P$ would appear in the translational flux $\bj_T$. However, this would not affect the resulting equation for $\Psi$. In fact, any advective linear velocity is allowed provided that the translational flux vanishes at infinity. A difficulty in the marginalization would appear if the angular velocity $\bOmega$ or the rotational diffusivity $D_R$ depends on $\bx$. For linear flows as we consider here, the angular velocity is spatially uniform.

\subsection{Rotational Taylor dispersion theory}
It is cumbersome to work with the unit orientation vector $\bq$ in the consideration of the long-time rotational dispersion because $\bq$ is bounded to the unit sphere~\citep{kammerer1997,Michele2001,Mazza2006,Mazza2007,Hunter11}. From a micromechanical perspective in considering the stochastic trajectory of a particle, one needs to be able to track the unbounded or cumulative angular displacement. The particle orientation vector is constrained to rotate in the velocity-gradient plane~\citep{Leahy2013,leahy_koch_cohen_2015}. In two dimensions, the orientation vector can be parameterized as $\bq = \cos\phi \be_x + \sin\phi\be_y$, where $\be_x$ and $\be_y$ are the unit basis vectors of the Cartesian coordinate system $(x,y)$, and $\phi \in [0,2\pi)$ is the orientation angle. The cumulative angular displacement $\varphi$ that is not bounded to the interval $[0, 2\pi)$ can be defined via
\begin{equation}
\label{eq:global-local-2d}
    \varphi = 2 \pi j + \phi, 
\end{equation}
where $j \in \mathbb{Z}$. Conversely, the bounded orientation angle $\phi$ is $\varphi$ modulo $2\pi$. For a constant angular velocity $\bOmega = \Omega \be_z$ with $\be_z=\be_x\times \be_y$, we have $\varphi(t) - \varphi(0) = \int_0^t \Omega \odiff s =  \Omega t$, where $\varphi(0)$ is a reference value.

We remark that alternative methods exist to quantify the rotational dynamics. In particular, one may extract a long-time dispersion coefficient from an orientational correlation function as a function of time in a BD simulation of the orientational Langevin equation of motion~\citep{dhont1996introduction,zwanzig2001nonequilibrium,Leahy2013}. One can also directly keep track of the unbounded angular displacement $\varphi$ in a BD simulation and infer the long-time transport coefficients~\citep{kammerer1997,Michele2001,Mazza2006,Mazza2007,Hunter11}. Because our aim is to derive a GTD theory from a continuum (Smoluchowski) perspective, such methods are not pursued here. In \citet{leahy_koch_cohen_2015} a coordinate transformation is discovered and used to  map the orientational dynamics to a diffusion equation, which ultimately leads to a closed-form asymptotic solution to the long-time rotational diffusivity in the high shear rate limit.

In terms of the bounded orientation angle $\phi$, the Smoluchowski equation \eqref{eq:smol-gen} is written as 
\begin{equation}
\label{eq:smol-2d}
    \frac{\partial \Psi }{\partial t} + \frac{\partial }{\partial \phi}\left( \Omega(\phi) \Psi - D_R \frac{\partial \Psi }{\partial \phi}\right)=0,
\end{equation}
 where the angular velocity $ \Omega(\phi) $ depends on the orientation angle.  It is clear that \eqref{eq:smol-2d} remains unchanged in terms of the unbounded coordinate $\varphi$. Noting that $\Psi = \Psi(\varphi, t) = \Psi(j, \phi, t)$,  one can  rewrite $\Psi$ in terms of the sequence \{$\Psi_j(\phi, t)$,  $\forall j \in \mathbb{Z}$ \}. In other words, to locate the particle in the unbounded orientation space, one can first identify the cell index, $j$, in which the particle resides,  and then use the local angular position $\phi$ within this cell. In the language of GTD, $\phi$ is identified as the local coordinate whereas the cell index $j$ is the global coordinate~\citep{frankel1989foundations,brenner1993macrotransport}. In other words, the long-time diffusive dynamics holds only when the particles have traversed many cells.

Because the $\varphi$ space is unbounded, it is more convenient to work in Fourier space. In the following, we make use of the flux-averaging approach of Brady and coworkers~\citep{morris_brady_1996,zia_brady_2010,takatori2014swim,Takatori17,Burkholder17,PB2020}. We note that the original approach was developed for the transport of particles in unbounded domains where the global coordinate is continuous (e.g., the longitudinal coordinate along a flat channel). Recently, \citet{Takatori20} extended this approach to accommodate the transport and dispersion in an oscillating array of harmonic traps, where the global coordinate is the discrete cell index of the infinite lattice. (An equivalent real-space approach may be taken where one makes use of the method of moments; see, e.g.~\citealt{brenner1980dispersion,alonso2019transport}. ) In the current problem, we have a one-dimensional lattice of unit cells. Following \citet{Takatori20}, we introduce the semi-discrete Fourier transform \citep{trefethen2000spectral}
\begin{equation}
    \hat{f}(k) = 2\pi \sum_{j = -\infty }^\infty e^{- ik j 2\pi} f_j, 
\end{equation}
where $k$ is the wavenumber and $i$ ($i^2=-1$) is the imaginary unit. Note that the transform is from $j$ to $k$,  and the local  coordinate $\phi$ is unchanged. In Fourier space, the Smoluchowski equation becomes 
\begin{equation}
\label{eq:smol-2d-Fourier}
    \frac{\partial \hat{\Psi}}{\partial t} + \left(ik + \frac{\partial }{\partial \phi}\right)\left[ \Omega \hat{\Psi} - D_R \left(ik + \frac{\partial }{\partial \phi} \right)\hat{\Psi}\right]=0,
\end{equation}
where $\hat{\Psi} = \hat{\Psi}(k, \phi, t)$ is the Fourier transform of $\Psi$. We note that in \eqref{eq:smol-2d-Fourier} the local gradient is written in real space whereas the global gradient appears in powers of $k$. The cell-averaged distribution, 
\begin{equation}
    \bavg{\hat{\Psi}}(k,t) = \frac{1}{2\pi}\int_0^{2\pi} \hat{\Psi}(k, \phi, t) \odiff\phi, 
\end{equation}
satisfies 
\begin{equation}
\label{eq:Phi-avg-eq}
    \frac{ \partial \bavg{\hat{\Psi}} }{\partial t} + ik \bavg{\Omega \hat{\Psi}} + D_R k^2 \bavg{\hat{\Psi}}=0, 
\end{equation}
which is obtained by averaging \eqref{eq:smol-2d-Fourier}. In writing \eqref{eq:Phi-avg-eq} we have invoked the periodic boundary condition on the local coordinate $\phi$~\citep{Takatori20}. One can relate $\hat{\Psi}$ to its average by defining the structure function $\hat{G}$ such that $\hat{\Psi}(k,\phi, t) = \bavg{\hat{\Psi}}(k,t) \hat{G}(k,\phi,t)$. The structure function is normalized, i.e.
\begin{equation}
     \bavg{\hat{G}} =1. 
\end{equation}

To derive an effective advection-diffusion equation for $\bavg{\hat{\Psi}}$, we first take a small wavenumber expansion of $\hat{G}$~\citep{morris_brady_1996,zia_brady_2010}, giving 
\begin{equation}
\label{eq:small-k-2d}
    \hat{G}(k,\phi,t) = g(\phi, t) + ik \, b(\phi, t)+O(k^2), 
\end{equation}
where $g$ is the average (zero wavenumber) field and $b$ is the displacement field. Inserting the expansion $\bavg{\Omega \hat{\Psi}} = \bavg{\hat{\Psi}} \bavg{\Omega (g + ik\, b)} +O(k^2)$ into \eqref{eq:Phi-avg-eq}, we obtain 
\begin{subequations}
\label{eq:macrotransport}
    \begin{equation}
    \label{eq:adv-diffusion-2d-eq}
    \frac{ \partial \bavg{\hat{\Psi}} }{\partial t} + ik \Omega^\text{eff} \bavg{\hat{\Psi}} + k^2 D^\text{eff}  \bavg{\hat{\Psi}}=0,
\end{equation}
\begin{equation}
\label{eq:Omega-D-eff-2d}
    \Omega^\text{eff} = \bavg{\Omega g}, \quad D^\text{eff} = D_R - \bavg{\Omega b}. 
\end{equation}
\end{subequations}
Note that \eqref{eq:adv-diffusion-2d-eq} is an effective advection-diffusion equation in Fourier space, with the effective rotational drift and rotational dispersion coefficient given in \eqref{eq:Omega-D-eff-2d}. 

Subtracting \eqref{eq:Phi-avg-eq} multiplied by $\hat{G}$ from \eqref{eq:smol-2d-Fourier}, we obtain 
\begin{equation}
\label{eq:hG-eq-2d}
    \frac{\partial \hG}{\partial t}+ ik \left[ \left( \Omega - \frac{\bavg{\Omega \hPhi }}{\bavg{\hPhi}}\right)\hG -D_R \frac{\partial \hG }{\partial \phi}\right] + \frac{\partial}{\partial \phi}\left[ \Omega \hG - D_R \left(ik +\frac{\partial }{\partial \phi}\right)\hG\right]=0. 
\end{equation}
Inserting the expansion \eqref{eq:small-k-2d} into \eqref{eq:hG-eq-2d}, at $O(1)$ we obtain 
\begin{equation}
\label{eq:g}
    \frac{\partial g}{\partial t} + \frac{\partial}{\partial \phi}\left( \Omega g - D_R \frac{\partial g}{\partial \phi}\right)=0. 
\end{equation}
At $O(k)$, we have
\begin{equation}
\label{eq:beq}
    \frac{\partial b}{\partial t} + \frac{\partial}{\partial \phi}\left( \Omega b - D_R \frac{\partial b}{\partial \phi}\right)=2 D_R \frac{\partial g}{\partial \phi} + \left( \Omega^\text{eff} - \Omega\right)g. 
\end{equation}
The average displacement field vanishes: $\avg{b}=0$. Equations \eqref{eq:macrotransport}, \eqref{eq:g} and \eqref{eq:beq} are the main results of this paper.

It follows that for a particle undergoing a steady rotation ($\Omega=const.$), $g=1$ and $b=0$, which implies that $\Omega^\mathrm{eff}=\Omega$ and $D^\mathrm{eff} = D_R$. As a result, a non-uniform or $\bq$-dependent angular velocity is required to achieve a long-time dispersivity that is potentially different from the bare diffusivity $D_R$. To calculate the average drift, one needs to solve \eqref{eq:g} and then take the average of $\Omega g$. With the solution of \eqref{eq:g}, one can solve \eqref{eq:beq} and then use \eqref{eq:Omega-D-eff-2d} to calculate the dispersion coefficient.

\section{Results}
\label{sec:results}
\begin{figure}
  \centerline{\includegraphics[width=5in]{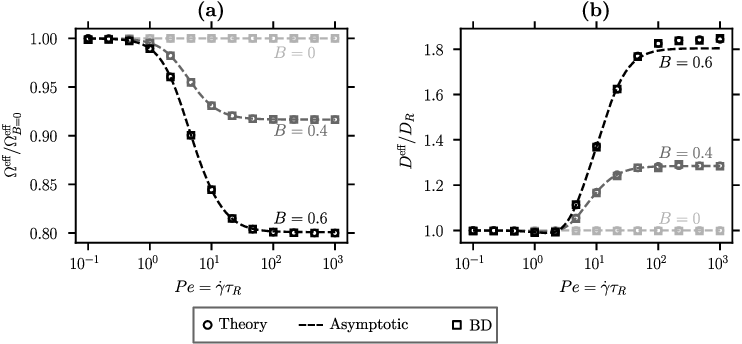}}
  \caption{(a) Plots of the average angular drift velocity scaled by the sphere result as a function of $Pe$ for different values of $B$. (b) Plots of the non-dimensional effective long-time dispersion coefficient  as a function of $Pe$ for different values of $B$. For spheres ($B=0$), the dispersion coefficient is not affected by the flow. For non-spherical particles, shear-enhanced dispersion is observed. }
\label{fig:steady-shear-2d}
\end{figure}

\subsection{Simple shear}
\label{subsec:simple-shear}
Consider the simple shear flow given by $\bu = \dot{\gamma} y \be_x$, where $\be_x$ is the unit basis vector in the $x$ direction of the Cartesian coordinate system $(x,y,z)$, $\dot{\gamma}$ is the shear rate. The problem is non-dimensionalized using the time scale $\tau_R = 1/D_R$. Two dimensionless groups dictate the behaviour of the problem. The first is a P\'eclet number, $Pe = \dot{\gamma}\tau_R$, which compares the time scale of the flow to that of rotational diffusion.  The second parameter is the Bretherton constant that characterizes the aspect ratio of the spheroid~\citep{bretherton1962motion}.  The non-dimensional (scaled by $\tau_R$) angular velocity is 
\begin{equation}
    \Omega \tau_R = -\frac{1}{2}Pe \left[ 1 - B \cos(2\phi) \right]. 
\end{equation}

For spherical particles, $B =0$, and the angular velocity is a constant, $\Omega\tau_R = -Pe/2$. In this case, $g = 1$ and $\Omega^\mathrm{eff} = \Omega$. From \eqref{eq:beq}, we readily obtain $b =0$ and $D^\mathrm{eff} = D_R$. Because the angular velocity is a constant, the average drift is simply the angular velocity of the flow and the flow does not affect the dispersion coefficient. Similar to the classical Taylor dispersion in which a spatially non-uniform advection is present, an orientation-dependent angular velocity is required to have potentially flow-enhanced dispersion~\citep{Leahy2013,leahy_koch_cohen_2015}.

To probe the effect of non-uniform angular velocity on the long-time drift and dispersion, we seek a regular series solution by writing 
\begin{equation}
\label{eq:B-expansion-def}
    g= \sum_{n=0}^\infty B^n g_n \quad \mathrm{and}\quad b = \sum_{n=0}^\infty B^n b_n. 
\end{equation}
The resulting drift and dispersion coefficient are written as, respectively, 
\begin{equation}
\label{eq:Omega-Deff-expansion-def}
    \Omega^\mathrm{eff} \tau_R = \sum_{n=0}^\infty B^n \Omega^\mathrm{eff}_n \quad \mathrm{and}\quad \frac{D^\mathrm{eff}}{D_R} = \sum_{n=0}^\infty B^n D^\mathrm{eff}_n. 
\end{equation}
We calculate the series solution up to $O(B^6)$ in Appendix \ref{sec:appendix-simple-shear}. The drift terms are given by 
\begin{subequations}
 \begin{eqnarray}
    \Omega^\mathrm{eff}_0 &=& -\frac{1}{2}Pe, \quad \Omega^\mathrm{eff}_2 = \frac{ Pe^3}{4 Pe^2+64}, \quad \Omega^\mathrm{eff}_4=  \frac{ Pe^5 \left(Pe^2-80\right)}{16 \left(Pe^2+16\right)^2 \left(Pe^2+64\right)},\\
      \Omega^\mathrm{eff}_6 &=& \frac{ (Pe-4) Pe^7 (Pe+4) \left(Pe^2-368\right)}{32 \left(Pe^2+16\right)^3 \left(Pe^2+64\right) \left(Pe^2+144\right)},
\end{eqnarray}   
\end{subequations}
where the odd terms are zero. For the dispersion coefficient, we obtain 
\begin{subequations}
 \begin{eqnarray}
    &&D^\mathrm{eff}_0 =1, \quad D^\mathrm{eff}_2 = \frac{Pe^2 \left(3 Pe^2-16\right)}{2 \left(Pe^2+16\right)^2}, \\ 
    &&D^\mathrm{eff}_4 = \frac{Pe^4 \left(3 Pe^6-124 Pe^4-12992 Pe^2+20480\right)}{2 \left(Pe^2+16\right)^3 \left(Pe^2+64\right)^2},\\
      &&D^\mathrm{eff}_6 = \frac{3C\,Pe^6 }{2 \left(Pe^2+16\right)^4 \left(Pe^4+208 Pe^2+9216\right)^2},
\end{eqnarray}   
\end{subequations}
where $C = -36175872 + 51011584 Pe^2 - 919040 Pe^4 - 47328 Pe^6 - 170 Pe^8 + Pe^{10}$ and odd terms vanish.

In figure \ref{fig:steady-shear-2d} we plot the average drift scaled by the drift of a sphere and the dispersion coefficient as a function of $Pe$ for several values of $B$. The scaled drift is shown in figure \ref{fig:steady-shear-2d}(a) and the non-dimensional dispersion coefficient is plotted in figure \ref{fig:steady-shear-2d}(b). The truncated series solution is shown by dashed lines. The circles in figure \ref{fig:steady-shear-2d} are results obtained by solving \eqref{eq:g} and \eqref{eq:beq} at steady state using a Fourier collocation method. The squares are from BD simulations of the orientational Langevin equation. In two dimensions, the Langevin equation (dimensional) is written as $\odiff \varphi/dt = \Omega + \sqrt{2D_R}\xi$, where $\xi$ is a white-noise process satisfying $\overline{\xi(t)}=0$ and $\overline{\xi(t)\xi(t^\prime)} = \delta(t-t^\prime)$. Here, the overhead bar denotes an ensemble average and $\delta$ is the delta function. We remark that in the Langevin equation, the unbounded angular coordinate $\varphi$ is used in order to calculate the mean and mean-squared angular displacements, from which the drift and dispersion coefficient can be obtained.  The full numerical solutions (circles) of \eqref{eq:g} and \eqref{eq:beq}  agree with the results from BD (squares), which validates our theory.

For spheres, $B=0$, and the drift is equal to the constant angular velocity.  As $B$ increases, the drift decreases compared with that of the sphere because the alignment term due to the rate of strain becomes more important. In the limit $Pe \to 0$, the drift of non-spherical particles approaches that of spheres, $\Omega^\mathrm{eff}/\Omega^\mathrm{eff}_{B=0} \to 1$.  The reduction of the scaled drift occurs at non-zero $Pe$ and is most prominent for large $Pe$ where the scaled drift asymptotes to a plateau as $Pe \to \infty$. For non-spherical particles, we observe a shear-enhanced angular dispersion as shown in figure \ref{fig:steady-shear-2d}(b). Similar to the classical Taylor dispersion in linear position space, the enhanced angular dispersion is a result of the coupling between non-uniform advection and diffusion. For spheres, the angular velocity is constant and no shear-enhanced dispersion is observed. When $B >0$, the dispersion coefficient increases monotonically as a function of $Pe$ until it asymptotes to a plateau at large $Pe$. In dimensional terms, we have $D^\mathrm{eff}/D_R = O(Pe^0)$ as $Pe \to \infty$, which is different from the classical Taylor dispersion of Brownian solutes in Poiseuille flow where $D^\mathrm{eff}/D_T \sim Pe^2$ as $Pe \to \infty$. For dispersion in position space, $D_T$ is the translational diffusivity and $Pe = U L/D_T$ with $U$ the characteristic fluid velocity and $L$ the characteristic width of the channel. We further note that Brownian particles in unbounded shear flows exhibit anomalous diffusion in position space~\citep{san1979brownian,foister1980diffusion,Krishnan,katayama1996brownian,Orihara11,takikawa2012diffusion}.

\begin{figure}
  \centerline{\includegraphics[width=5in]{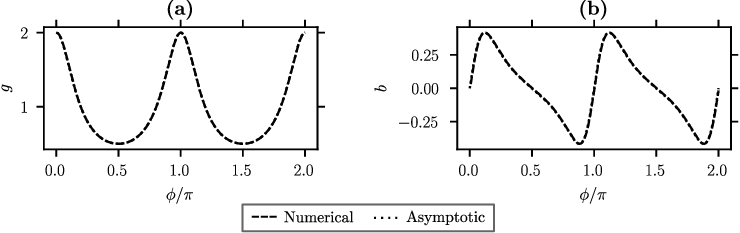}}
  \caption{ Plots of numerical solutions (dashed) at $Pe=1000$ and the leading-order asymptotic solutions (dotted) at large $Pe$ for (a) the average field and (b) the displacement field. In the plots, $B=0.6$.  The numerical and asymptotic solutions have excellent agreement and cannot be distinguished visually. }
\label{fig:g-b-largePe}
\end{figure}

\begin{figure}
  \centerline{\includegraphics[width=5in]{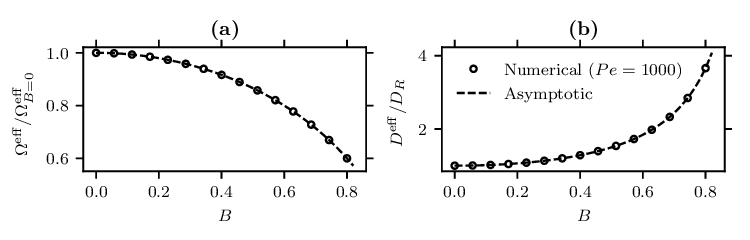}}
  \caption{ Plots of numerical solutions (circles) at $Pe=1000$  and the leading-order asymptotic solutions (dashed) at large $Pe$ for (a) the scaled drift and (b) the dispersion coefficient. }
\label{fig:Omega-Deff-largePe}
\end{figure}

To understand the behaviour of the system at large $Pe$, we consider a perturbation expansion in powers of $1/Pe$, 
\begin{subequations}
    \begin{eqnarray}
    \label{eq:g-series-large-Pe}
        g &=&g_0 + \frac{1}{Pe}g_1 + \frac{1}{Pe^2}g_2+\cdots, \\ 
        \label{eq:b-series-large-Pe}
        b &=&b_0 + \frac{1}{Pe}b_1 + \frac{1}{Pe^2}b_2 +\cdots.
\end{eqnarray}
\end{subequations}
Inserting the expansion \eqref{eq:g-series-large-Pe} into \eqref{eq:g}, one can solve the resulting equations order by order. We derive after some algebra that 
\begin{subequations}
\label{eq:g-solution-large-Pe}
    \begin{eqnarray}
    \label{eq:g0-large-Pe}
        g_0 &=&  \frac{\sqrt{1 - B^2}}{1 - B \cos(2\phi)},\\
        g_1 &= &\frac{4B \sqrt{1-B^2} \sin(2 \phi) }{\left[ 1-B\cos(2\phi)\right]^3}. 
    \end{eqnarray}
\end{subequations}
We note that in obtaining the preceding solutions, the conservation conditions $\avg{g_0}=1$ and $\avg{g_1}=0$ are enforced.  From \eqref{eq:g-solution-large-Pe}, we obtain
\begin{equation}
\label{eq:large-Pe-Omegaeff}
    \frac{\Omega^\mathrm{eff}}{\Omega^\mathrm{eff}_{B=0}} = \sqrt{1-B^2} +O(1/Pe^2)\quad\mathrm{as}\quad Pe \to \infty.
\end{equation}
As $Pe \to \infty$, the scaled drift approaches a finite value that depends on $B$. Due to symmetry, the function $g_1$ does not contribute to the drift. The correction  is $O(1/Pe^2)$, which comes from $g_2$ in the expansion. In figure \ref{fig:g-b-largePe}(a) the leading-order asymptotic solution $g_0$ is plotted as a function of $\phi$ for $B=0.6$. The full numerical solution of \eqref{eq:g} at $Pe=1000$ and $B=0.6$ is also shown in figure \ref{fig:g-b-largePe}(a). In figure \ref{fig:Omega-Deff-largePe}(a) we plot the leading-order drift given in \eqref{eq:large-Pe-Omegaeff} as a function of $B$ and the numerical solution of the macrotransport equations for $Pe=1000$. Good agreement between the asymptotic and numerical solutions is observed.

Substituting \eqref{eq:b-series-large-Pe} and \eqref{eq:g-solution-large-Pe} into \eqref{eq:beq}, we obtain at $O(1)$
\begin{equation}
     \frac{\partial}{\partial \phi}\left[ -\frac{1}{2}\left(1-B\cos(2\phi)\right)b_0\right] = \left[ -\frac{1}{2} \sqrt{1-B^2} +\frac{1}{2}\left(1-B\cos(2\phi)\right) \right]g_0,
\end{equation}
where we have used \eqref{eq:large-Pe-Omegaeff}.  Defining 
\begin{equation}
\label{eq:tilde-b0-def}
    \tilde{b}_0(\phi) = \frac{\sqrt{1-B^2}}{1-B\cos(2\phi)} \left[ \arctan\left(\frac{(1+B) \tan \phi }{\sqrt{1-B^2}}\right) - \phi\right], \quad 0 \leq \phi  \leq  \pi/2, 
\end{equation}
one can write the solution at $O(1)$ as 
\begin{equation}
b_0(\phi) = 
    \begin{cases}
        \tilde{b}_0(\phi) &  0 \leq \phi \leq \pi/2, \\ 
         - \tilde{b}_0(\pi-\phi) &\pi/2 \leq  \phi \leq  \pi, \\ 
         \tilde{b}_0(\phi - \pi) & \pi \leq \phi \leq 3\pi/2,\\
         - \tilde{b}_0(2\pi-\phi) &3\pi/2 \leq  \phi \leq  2\pi. 
    \end{cases}
\end{equation}
In figure \ref{fig:g-b-largePe}(b) the leading-order asymptotic solution $b_0$ is plotted as a function of $\phi$ for $B=0.6$. The full numerical solution of \eqref{eq:beq} at $Pe=1000$ and $B=0.6$ is also shown in figure \ref{fig:g-b-largePe}(b). Good agreement between the asymptotic and numerical solutions is observed. Because of the symmetry of $\Omega$ and $b_0$, the average $\avg{\Omega b_0}$ vanishes.

\begin{figure}
  \centerline{\includegraphics[width=5in]{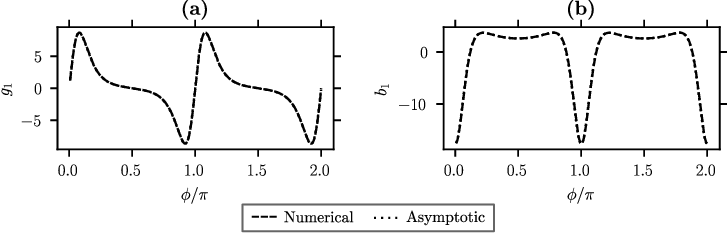}}
  \caption{ Plots of numerical solutions (dashed) at $Pe=1000$ and the $O(1/Pe)$ asymptotic solutions (dotted) at large $Pe$ for (a) the average field and (b) the displacement field. In the plots, $B=0.6$. The numerical approximation to $g_1$ is obtained via $g_1^\mathrm{num} =Pe\left( g^\mathrm{num} - g_0\right)$, where the superscript `num' denotes a numerical solution and $g_0$ is given in \eqref{eq:g0-large-Pe}. Similarly,  $b_1^\mathrm{num} =Pe\left( b^\mathrm{num} - b_0\right)$. The numerical and asymptotic solutions have excellent agreement and cannot be distinguished visually. }
\label{fig:g1-b1-largePe}
\end{figure}

To obtain the first non-zero term of shear-induced dispersion,  we therefore need to consider the $O(1/Pe)$ solution to $b$. At $O(1/Pe)$, the displacement field equation is given by 
\begin{equation}
\label{eq:b1-eq-largePe}
    \frac{\partial }{\partial \phi}\left[  -\frac{1}{2}[1-B \cos(2\phi)] b_1 - \frac{\partial b_0}{\partial \phi} \right] = 2 \frac{\partial g_0}{\partial \phi}. 
\end{equation}
Similar to \eqref{eq:tilde-b0-def}, we define
\begin{align}
\label{eq:tilde-b1-def}
     \tilde{b}_1(\phi) = \frac{4 B \sqrt{1-B^2} \sin (2 \phi )}{\left[1-B \cos (2 \phi )\right]^3} \left[ \arctan\left(\frac{(1+B) \tan \phi }{\sqrt{1-B^2}}\right) - \phi\right] +\frac{3 B^2-3}{\left[1-B \cos (2 \phi )\right]^3},
\end{align}
which is valid for $\phi \in [0, \pi/2]$. Using \eqref{eq:tilde-b1-def}, a particular solution to $b_1$ is constructed as 
\begin{equation}
b_1^p(\phi) = 
    \begin{cases}
        \tilde{b}_1(\phi) &  0 \leq \phi \leq \pi/2, \\ 
          \tilde{b}_1(\pi-\phi) &\pi/2 \leq  \phi \leq  \pi, \\ 
         \tilde{b}_1(\phi - \pi) & \pi \leq \phi \leq 3\pi/2,\\
          \tilde{b}_1(2\pi-\phi) &3\pi/2 \leq  \phi \leq  2\pi. 
    \end{cases}
\end{equation}
The full solution can be written as 
\begin{equation}
\label{eq:b1-full-sol}
    b_1(\phi) = \frac{B^2+\sqrt{1-B^2}+2}{1-B^2} \frac{1}{1-B\cos(2\phi)} + b_1^p(\phi), 
\end{equation}
where the first term on the right-hand side is the homogeneous solution. To compare the $O(1)$ asymptotic solutions with numerical solutions, we first construct a hybrid approximation to $g_1$, given by $g_1^\mathrm{num} =Pe\left( g^\mathrm{num} - g_0\right)$, where the superscript `num' denotes the numerical solution. In figure \ref{fig:g1-b1-largePe}(a), $g_1^\mathrm{num}$ is compared with the asymptotic solution $g_1$, and good agreement is observed. Similarly, we define $b_1^\mathrm{num} =Pe\left( b^\mathrm{num} - b_0\right)$. The results for $b_1^\mathrm{num}$ and $b_1$ are plotted in figure \ref{fig:g1-b1-largePe}(b).

Using \eqref{eq:b1-full-sol} and the expansion
\begin{equation}
    D^\mathrm{eff} = D_0^\mathrm{eff} + \frac{1}{Pe}D_1^\mathrm{eff}+\cdots, 
\end{equation}
we obtain 
\begin{equation}
\label{eq:Deff-large-Pe-shear}
    \frac{D_0^\mathrm{eff}}{D_R} = 1 + \Bavg{\frac{1}{2}[1-B\cos(2\phi)] b_1} = \frac{2+B^2}{2\left(1-B^2\right)}. 
\end{equation}
In figure \ref{fig:Omega-Deff-largePe}(b) we compare the asymptotic solution given in \eqref{eq:Deff-large-Pe-shear} with the full numerical solution of the macrotransport equations for $Pe=1000$. We note that \eqref{eq:Deff-large-Pe-shear} agrees with the result obtained by \citet{leahy_koch_cohen_2015} where a coordinate transformation is employed to map the orientational dynamics to a diffusion equation in the large-$Pe$ limit. In contrast to their approach, the current theory follows closely the GTD framework and allows us to calculate both the average drift and effective dispersion for arbitrary P\'eclet numbers in generic linear flows.

\subsection{Extensional flow}
\label{subsec:extensional}
\begin{figure}
  \centerline{\includegraphics[width=3in]{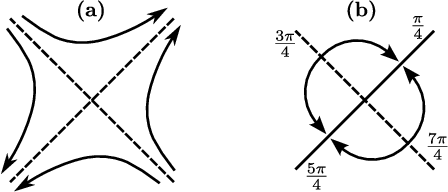}}
  \caption{ (a) Schematic of the extensional flow. (b) Schematic of the direction of rotation due to the extensional flow.} 
\label{fig:extensional-flow}
\end{figure}

As another case study, we now consider an extensional flow where the angular velocity is  $\bOmega = Pe\,B \cos(2\phi) \be_z$. The extensional flow tends to align the particle orientation with the extensional axis (see figure~\ref{fig:extensional-flow}). The particle has a non-zero angular velocity when the orientation deviates from the extensional axis. As shown in figure~\ref{fig:extensional-flow}(b), the direction of rotation depends on the particle orientation relative to the extensional axis. 

The average field in the extensional flow is governed by 
\begin{equation}
\label{eq:extension-2d-g-eq}
    \frac{\partial }{\partial \phi}\left(B Pe \cos(2\phi) g -  \frac{\partial g}{\partial \phi}\right)=0, 
\end{equation}
Because the particle can align with both directions of the extensional axis ($\phi=\pi/4$ and $5\pi/4$ in figure~\ref{fig:extensional-flow}(b)), the average field has a periodicity of $\pi$, $g(\phi+\pi) = g(\phi)$. Defining $\phi^\prime = \phi-\pi/4$, we consider $g$ in the interval $\phi^\prime \in [-\pi/2, \pi/2]$. In terms of the shifted variable $\phi^\prime$, we have 
\begin{equation}
\label{eq:extension-2d-g-shifted}
    \frac{\partial }{\partial \phi^\prime}\left(B Pe \sin(2\phi^\prime) g -  \frac{\partial g}{\partial \phi^\prime}\right)=0,  \quad \phi^\prime  \in \left[-\frac{\pi}{2}, \frac{\pi}{2} \right].
\end{equation}
From \eqref{eq:extension-2d-g-shifted}, we see that $g$ is an even function of $\phi^\prime$; this can also be understood by examining figure \ref{fig:extensional-flow}(b). The average angular drift $\avg{\Omega g}$ vanishes because $\Omega g$ is an odd function of $\phi^\prime$ in the interval $\phi^\prime \in [-\pi/2, \pi/2]$. 

Integrating \eqref{eq:extension-2d-g-eq} once, we obtain 
\begin{equation}
    B Pe \cos(2\phi) g -  \frac{\partial g}{\partial \phi}= C_1. 
\end{equation}
Averaging the above equation over one period, we have $\avg{\Omega g} = C_1$. Because $\avg{\Omega g}=0$ as determined from symmetry, we must have $C_1=0$. With this, we obtain 
\begin{equation}
    g(\phi) = \frac{1}{A_1}\exp\left(\frac{1}{2}B\,Pe \sin(2\phi)\right), \quad \phi \in [0, 2\pi],
\end{equation}
where $A_1$ is determined from $\avg{g}=1$.

The displacement field in the extensional flow satisfies 
\begin{equation}
\label{eq:b-eq-extension}
     \frac{\partial }{\partial \phi}\left(B Pe \cos(2\phi) b -  \frac{\partial b}{\partial \phi}\right)=2 \frac{\partial g}{\partial \phi} - B\, Pe \cos(2\phi) g = \frac{\partial g}{\partial \phi}. 
\end{equation}
Because \eqref{eq:b-eq-extension} does not admit a simple closed-form analytic solution, we instead seek pertubative solutions in the following two limits: (1) $Pe\,B\ll1$ and (2) $Pe\, B \gg 1$.

In the small $Pe\, B$ limit, we write
\begin{subequations}
    \begin{align}
        g(\phi) &= 1 + Pe\,B g_1(\phi) + \cdots, \\ 
        b(\phi) &= 0 + Pe\,B b_1(\phi) + \cdots,
    \end{align}
\end{subequations}
where 
\begin{subequations}
    \begin{align}
        g_1= \frac{1}{2}\sin(2\phi), \quad \mathrm{and}\quad b_1 =\frac{1}{4} \cos (2 \phi ). 
    \end{align}
\end{subequations}
From this, we determine the dispersion coefficient as 
\begin{equation}
\label{eq:Deff-extension-small-PeB}
    \frac{D^\mathrm{eff}}{D_R} = 1 - (Pe\,B)^2\bavg{\cos(2\phi)b_1}+O\left(Pe^4B^4\right) = 1 - \frac{1}{8}Pe^2B^2+O\left(Pe^4B^4\right). 
\end{equation}

\begin{figure}
  \centerline{\includegraphics[width=3in]{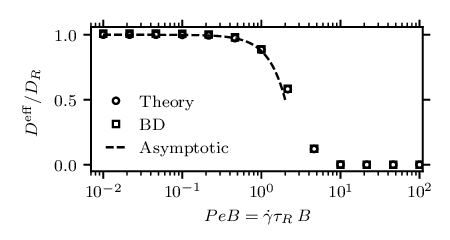}}
  \caption{ Plots of the effective dispersion coefficient as a function of $Pe\,B$ in the extensional flow. Circles are results obtained from the numerical solution of the Taylor dispersion theory, squares are results calculated from BD simulations and the dashed line denotes the leading-order asymptotic solution for small $Pe\,B$ given in  \eqref{eq:Deff-extension-small-PeB}.  } 
\label{fig:extensional-Deff}
\end{figure}

In figure~\ref{fig:extensional-Deff} we plot the effective dispersion coefficient as a function of $Pe\,B$ in the extensional flow. The leading-order asymptotic solution for small $Pe\,B$ in \eqref{eq:Deff-extension-small-PeB} is denoted by  the dashed line. In the small $Pe\,B$ regime, the asymptotic solution agrees well with both the full numerical solution (circles) of the macrotransport equations and results obtained from BD simulations (squares). In the extensional flow, rotational dispersion is hindered and the dispersion coefficient vanishes in the large $Pe\,B$ limit. Because the extensional flow acts to align the particle orientation with the extensional axis, in the strong flow limit random reorientations are suppressed.

We now consider the probability distributions in the strong flow limit characterized by $\epsilon = 1/(Pe\,B) \ll 1$. In the limit $Pe\,B\to \infty$ or $\epsilon \to 0$, the orientational distribution becomes a delta function localized at $\phi_0 = \pi/4$ due to strong alignment. For strong flow, the particle orientation is closely aligned with the extensional axis, which implies the existence of a boundary layer near $\phi_0 = \pi/4$. A dominant balance reveals that the boundary layer thickness is $O(\sqrt{\epsilon})$. Introducing the stretched coordinate $\xi = (\phi- \phi_0)/\sqrt{\epsilon}$, we rewrite \eqref{eq:extension-2d-g-eq} in the boundary layer as 
\begin{equation}
    \frac{\partial }{\partial \xi}\left( \cos \left(2 \phi_0 + 2\epsilon^{1/2} \xi\right) \tilde{g} - \sqrt{\epsilon} \frac{\partial \tilde{g}}{\partial \xi}\right)=0,
\end{equation}
where $\tilde{g}(\xi) = g(\phi)$. Noting that $\cos(2\phi_0)=0$ and $\cos \left(2 \phi_0 + 2\epsilon^{1/2} \xi\right)=-2 \xi  \sqrt{\epsilon }+\frac{4}{3} \xi ^3 \epsilon ^{3/2}-\frac{4}{15} \xi ^5 \epsilon ^{5/2} +O(\epsilon^{7/2})$, we expect an expansion of the form 
\begin{equation}
\label{eq:g-expansion-extension-LargePe}
    \tilde{g} = \frac{1}{\sqrt{\epsilon}}\tilde{g}_0 + \cdots,
\end{equation}
where the leading-order term is $O(1/\sqrt{\epsilon})$ due to the conservation of probability. In the bulk (outside the boundary layer), the orientational distribution is zero to algebraic orders of $\sqrt{\epsilon}$.

\begin{figure}
  \centerline{\includegraphics[width=5in]{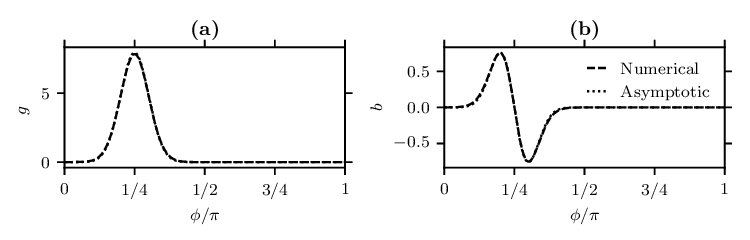}}
  \caption{ (a) Plots of the numerical solution of $g$ (dashed) at $Pe\,B=20$ and the leading-order asymptotic solution (dotted) given in \eqref{eq:large-Pe-extension-composite}. (b) Plots of the numerical solution of $b$ (dashed) at $Pe\,B=20$ and the leading-order asymptotic solution (dotted) given in \eqref{eq:b-0-extension-largePe}. } 
\label{fig:extensional-flow-largePe-g}
\end{figure}

The leading-order orientational distribution is governed by 
\begin{equation}
\label{eq:g0-tilde-BL}
    \frac{\partial }{\partial \xi}\left( 2 \xi \tilde{g}_0 + \frac{\partial \tilde{g}_0}{\partial \xi} \right)=0,
\end{equation}
which admits a solution of the form $\tilde{g}_0 = A_0 e^{-\xi^2}$, where $A_0$ remains to be determined. In writing the solution, we have made use of the matching condition $\tilde{g}\to 0$ as $\xi\to \infty$. From the conservation, $\avg{g}=1$, we obtain $A_0 = \sqrt{\pi}$. Due to symmetry, the leading-order solution is valid as $\phi$ approaches $\phi_0$ from either side. We then construct a leading-order composite solution as  
\begin{equation}
\label{eq:large-Pe-extension-composite}
g = \sqrt{\frac{\pi}{\epsilon}} \exp\left(  - \frac{\left(\phi-\phi_0\right)^2}{\epsilon}\right) +\cdots,
\end{equation}
where the dots denote higher-order corrections.  In figure \ref{fig:extensional-flow-largePe-g}(a) we compare the leading-order solution in \eqref{eq:large-Pe-extension-composite} with the numerical solution of the full equation for $Pe\,B=20$. We  note that indeed the leading-order solution in \eqref{eq:large-Pe-extension-composite} approaches a delta function upon appropriate scaling in the limit $\epsilon \to 0$.

We now proceed to analyse the displacement field, which in the boundary layer is expanded as 
\begin{equation}
    \tilde{b} = \tilde{b}_0 +\cdots,
\end{equation}
where we remark that $\tilde{b}$ is $O(1)$ at leading order. At leading order, we obtain 
\begin{equation}
    - \frac{\partial }{\partial \xi}\left( 2 \xi \tilde{b}_0 + \frac{\partial \tilde{b}_0}{\partial \xi} \right)= 2\xi \tilde{g}_0+ 2\frac{\partial \tilde{g}_0}{\partial \xi}.
\end{equation}
The equation is solved by 
\begin{equation}
    \tilde{b}_0=\left(C_0 - \sqrt{\pi} \xi\right)e^{-\xi ^2}.
\end{equation}
Using the normalization condition $\avg{b}=0$, we must have $\int_{-\infty}^\infty \tilde{b}_0\odiff\xi=0$, which gives $C_0=0$. In terms of $\phi$, we may write 
\begin{equation}
\label{eq:b-0-extension-largePe}
    b_0(\phi) = -\sqrt{\pi} \frac{\phi- \phi_0}{\sqrt{\epsilon}}\exp\left(  - \frac{\left(\phi-\phi_0\right)^2}{\epsilon}\right). 
\end{equation}
 In figure \ref{fig:extensional-flow-largePe-g}(b) we compare the leading-order solution of $b$ in \eqref{eq:b-0-extension-largePe} with the numerical solution of the full equation for $Pe\,B=20$.

 Taking the limit $\epsilon \to 0$, we obtain 
 \begin{equation}
     \frac{D^\mathrm{eff}}{D_R} =1 - \avg{\Omega b} \to  1- \frac{1}{\pi}\int_{-\infty}^\infty (-2) \xi \tilde{b}_0(\xi)  \odiff\xi =0. 
 \end{equation}
As a result, we have shown that the dispersion coefficient vanishes in the strong flow limit. This asymptotic result agrees with the full numerical solution and the BD simulation results (see figure~\ref{fig:extensional-Deff}).

\subsection{General linear flows}
We now consider a general two-dimensional linear flow of the form $\bu = \dot{\gamma}y \be_x + \alpha \dot{\gamma}x \be_y$, where $\alpha$ is a dimensionless parameter that controls the characteristics of the external flow. As a function of $\alpha$, the flow varies from purely rotational ($\alpha=-1$) to simple shear ($\alpha=0$) and extensional flow ($\alpha=1$). From \eqref{eq:Jeffery}, we obtain the angular velocity of the particle as 
\begin{equation}
    \label{eq:Omega-general}
    \Omega \tau_R =  \frac{1}{2}Pe \left[ -1+\alpha +B\left(1+\alpha\right)\cos(2\phi)\right]. 
\end{equation}

\begin{figure}
    \centerline{\includegraphics[width=3in]{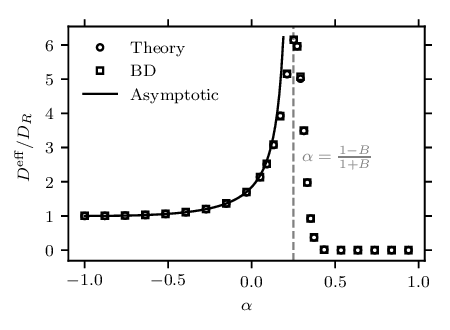}}
    \caption{ Plots of the dispersion coefficient as a function of $\alpha$ for $Pe=100$ and $B=0.6$. Circles are results from numerical solutions of the macrotransport equations. Squares are results from BD simulations. The vertical line is given by the equation $\alpha=(1-B)/(1+B)$. The solid line is the asymptotic solution in equation \eqref{eq:Deff-generic-largePe-regular}.} 
\label{fig:Deff-vs-alpha}
\end{figure}

To probe the effect of $\alpha$ on the long-time rotational dispersion coefficient, we solve the macrotransport equations \eqref{eq:g} and \eqref{eq:beq} numerically using a Fourier collocation method for $Pe=100$. In figure \ref{fig:Deff-vs-alpha} the numerical results are plotted as circles, and the results obtained from BD simulations are marked by squares. The dispersion coefficient exhibits a non-monotonic dependence on the flow parameter $\alpha$. The maximum of the dispersion coefficient is obtained when $\alpha = (1-B)/(1+B)$. For a general linear flow, the angular velocity is a combination of a constant part (purely rotational) and a varying part (extensional flow). For $\alpha=-1$, the flow is purely rotational, and we have $D^\mathrm{eff}=D_R$. As $\alpha$ increases, the varying part of the angular velocity gives enhanced dispersion. In the absence of the constant part ($\alpha=1$), particles becomes strongly aligned. As a result,  the dispersion is suppressed (see section \ref{subsec:extensional}). To achieve maximum enhancement in dispersion, the two terms need to balance, which occurs when $1-\alpha = B(1+\alpha)$ or $\alpha = (1-B)/(1+B)$. In figure \ref{fig:Deff-vs-alpha} this value of $\alpha$ is marked by a vertical line for $B=0.6$. On the right side of the vertical line, the extensional flow becomes dominant, and the dispersion coefficient decays to zero as $\alpha$ increases and approaches that of the extensional flow ($\alpha=1$).

\begin{figure}
    \centerline{\includegraphics[width=2.5in]{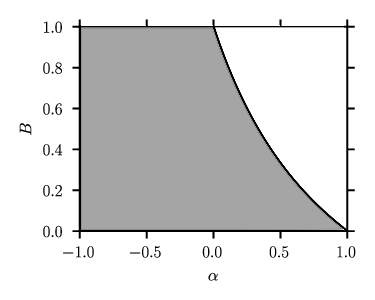}}
    \caption{ Plot of the parameter space $(\alpha, B)$. The solid line is given by $1-\alpha = B(1+\alpha)$. In the shaded region, we have $1-\alpha > B(1+\alpha)$.} 
\label{fig:singular_region}
\end{figure}

In the large-$Pe$ limit, we consider a perturbation expansion in powers of $1/Pe$, 
\begin{subequations}
    \begin{equation}
        g = g_0 + \frac{1}{Pe}g_1+\cdots,
    \end{equation}
    \begin{equation}
        b = b_0 + \frac{1}{Pe}b_1+\cdots. 
    \end{equation}
\end{subequations}
Following the notations and calculations in section \ref{subsec:simple-shear}, we obtain 
\begin{subequations}
\begin{equation}
\label{eq:g0-general}
g_0= \frac{A_1}{1-\alpha -B(1+\alpha)\cos(2\phi)},
\end{equation}
\begin{equation}
\label{eq:g1-general}
g_1 = \frac{4B(1+\alpha)A_1\sin(2\phi)}{\left[ 1-\alpha -B(1+\alpha)\cos(2\phi) \right]^3}, 
\end{equation}
\end{subequations}
where $A_1=\sqrt{(1-\alpha)^2-B^2(1+\alpha)^2}$. We note that \eqref{eq:g0-general} and \eqref{eq:g1-general} are only valid in the region $1-\alpha > B(1+\alpha)$, which is shown as the shaded region in figure \ref{fig:singular_region}. At the boundary, $1-\alpha = B(1+\alpha)$, the solutions in \eqref{eq:g0-general} and \eqref{eq:g1-general} becomes singular at $\phi_0$, where $\cos(2\phi_0)=1$ or $\phi_0\in \{0,\pi,2\pi\}$. One may similarly obtain the displacement field as 
\begin{equation}
\label{eq:b0-general}
\tilde{b}_0=A_1 \frac{\arctan\left(\frac{1-\alpha +B(1+\alpha)}{A_1}\tan\phi\right) -\phi}{1-\alpha - B(1+\alpha)\cos(2\phi)},
\end{equation}
\begin{equation}
\label{eq:b1-general}
\tilde{b}_1 = \frac{C_1\left[\arctan\left(\frac{(1-\alpha +B(1+\alpha))}{A_1}\tan\phi\right) -\phi\right]+C_2}{\left[1-\alpha - B(1+\alpha)\cos(2\phi)\right]^3}+ \frac{C_3}{1-\alpha - B(1+\alpha)\cos(2\phi)}, 
\end{equation}
where 
\begin{subequations}
\begin{equation}
C_1=4B(1+\alpha)A_1,\quad C_2=-3 A_1^2, 
\end{equation}
\begin{equation}
C_3=\frac{B^2(1+\alpha)^2+(1-\alpha)\left(2(1-\alpha)+A_1 \right)}{A_1^2}.
\end{equation}
\end{subequations}

The leading-order dispersion coefficient can then be shown to be 
\begin{equation}
\label{eq:Deff-generic-largePe-regular}
\frac{D_0^\mathrm{eff}}{D_R} = \frac{2(1-\alpha)^2+B^2(1+\alpha)^2}{2(1-\alpha)^2-2B^2(1+\alpha)^2}.
\end{equation}
For simple shear, $\alpha=0$, we recover equation \eqref{eq:Deff-large-Pe-shear}. In figure \ref{fig:Deff-vs-alpha}  the solution in \eqref{eq:Deff-generic-largePe-regular} for $B=0.6$ is plotted as a solid line. The asymptotic solution agrees well with both the numerical and BD results. As expected, the asymptotic solution deviates from the numerical or BD results as $\alpha$ approaches the singular boundary where $1-\alpha = B(1+\alpha)$. Furthermore, the solution of \eqref{eq:Deff-generic-largePe-regular} become singular when $1-\alpha = B(1+\alpha)$. 

In Appendix \ref{sec:appendix-generic-boundary}, using asymptotic theory, we analyse the macrotransport equations and characterize the long-time dispersion behaviour for the case of $1-\alpha = B(1+\alpha)$ in the strong flow limit. When $1-\alpha = B(1+\alpha)$, our analysis reveals that there exist boundary layers of thickness $O(\epsilon^{1/3})$ at $\phi = n \pi$, $n \in \mathbb{Z}$, where $\epsilon = 1/(Pe(1-\alpha))$. We show that the dispersion coefficient is $O(\epsilon^{-2/3})$  as $\epsilon \to 0$. For a fixed $\alpha$, this means that $D^\mathrm{eff}/D_R = O(Pe^{2/3})$.

Our results show that depending on the parameters $\alpha$ and $B$, the rotational dispersion coefficient exhibits distinctly different scaling in the large-$Pe$ limit. For $1-\alpha > B(1+\alpha)$, we have $D^\mathrm{eff}/D_R = O(1)$ as $Pe \to \infty$. On the curve where $1-\alpha = B(1+\alpha)$, $\alpha \neq 1$, $D^\mathrm{eff}/D_R =O(Pe^{2/3})$. Finally, for extensional flow $D^\mathrm{eff}/D_R \to 0$ as $Pe \to \infty$. We note that as a function of $\alpha$, the transition from the case of $1-\alpha =B(1+\alpha)$ to the vanishing dispersion at $\alpha=1$ occurs in a small region, which is shown by the rapid decay of $D^\mathrm{eff}$ in figure \ref{fig:Deff-vs-alpha}. Within this small region, a perturbation analysis for a small deviation from the curve $1-\alpha =B(1+\alpha)$ may be useful. Such an analysis is left for future considerations.

\section{Concluding remarks}
\label{sec:conclusion}
In this paper we have developed a GTD theory that describes the long-time rotational dispersion of a spheroidal particle in linear flows that is constrained to rotate in the velocity-gradient plane. As is standard for Taylor dispersion, the average drift and the effective dispersion are treated in a unified framework by leveraging a flux-averaging method in Fourier space. The results obtained from the continuum theory are corroborated by BD simulations of the orientational equation of motion. Using asymptotic analysis in the strong flow limit, we have shown that a simple shear enhances while an extensional flow hinders the long-time rotational dispersion. More specifically, we showed that $D^\mathrm{eff}/D_R=O(1)$  in simple shear and $D^\mathrm{eff}/D_R \to 0$ in extensional flow as $Pe \to \infty$ for a given non-zero $B$. These results reveal that the long-time rotational dispersion depends qualitatively on the characteristics of the background flow.

While we focused on the long-time dispersion in steady flows, our GTD theory applies equally well to time-periodic flows such as oscillatory shear. In oscillatory shear one needs to first obtain the long-time (time-dependent) solutions to the average and displacement fields. In addition to the cell average employed for steady linear flows, a time average over the oscillation period is needed to obtain the long-time transport coefficients. 

In linear flows as considered in this paper, the angular velocity of the particle due to the flow does not depend on the spatial position. This spatial uniformity allows us to consider the conservation equation in orientation space only. A difficulty would arise if one wishes to consider the rotational dispersion of a particle in flows in the presence of no-slip boundaries such as pressure-driven channel flows. In such a quadratic flow field, the angular velocity depends linearly on the transverse coordinate. Mathematically, the marginalization of $P(\bx, \bq, t)$ does not lead to a closed equation for $\Psi(\bq, t)$. In this case, one often needs to solve the full probability distribution $P(\bx, \bq, t)$ in order to calculate the long-time translational or rotational transport properties. As a concrete example, consider the rotational dynamics of a `point' Brownian particle in a planar Poiseuille flow. The Smoluchowski equation can be written as 
\begin{equation}
    \frac{\partial P}{\partial t} + \bnabla\bcdot(\bu P - D_T \bnabla P) + \bnabla_R \bcdot(\bOmega P - D_R \bnabla P)=0,
\end{equation}
where $\bOmega$ depends on $y$, which is the transverse coordinate.  Following the consideration given in section \ref{sec:formulation}, one can show that the average field is governed by 
\begin{equation}
    \frac{\partial }{\partial y}\left(-D_T \frac{\partial g}{\partial y} \right) + \frac{\partial}{\partial \phi}\left( \Omega(y) g -D_R \frac{\partial g}{\partial \phi}\right)=0.
\end{equation}
At the channel walls ($y=\pm H$), the no-flux condition, $-\partial g/(\partial y)=0$, is imposed. The normalization involves both spatial and orientational integrals, which is given by $\avg{\int_{-H}^H g(y, \phi) \odiff y}=1$. Similarly, one may write down the governing equation for the displacement field.

The rotational Taylor dispersion theory can  accommodate particles of different shapes such as chiral or ring-shaped particles~\citep{singh2013rigid}. In the presence of hydrodynamic translation-rotation coupling, one again needs to consider the dynamics in both position and orientation space. In the case of general particle shapes, much like the case of a Brownian particle in Poiseuille flow, one performs the marginalization to obtain net orientational distributions after solving the average and displacement fields as functions of both $\bx$ and $\bq$.

The rotational Taylor dispersion theory is developed in two dimensions where the particle rotates about a single axis. To accommodate particles that rotate in three-dimensional space, one may consider the rotational dynamics about perpendicular axes. For a particular chosen axis of rotation, one can then utilize the current theory to characterize the long-time dynamics. We note that measuring rotational diffusion about perpendicular axes in three dimensions has been used in experiments~\citep{Rogers12,D3SM01320K}.  To conclude, we also note that similar to classical Taylor dispersion, our theory is only valid at long times.

\section*{Funding}
This work is supported by the Faculty of Engineering at the University of Alberta. 

\section*{Declaration of interests}
 The author reports no conflict of interest.
 \section*{Author ORCID}
 Zhiwei Peng https://orcid.org/0000-0002-9486-2837

\appendix
\section{Asymptotic solution in simple shear}
\label{sec:appendix-simple-shear}
In this appendix we discuss the asymptotic solution for nearly spherical particles in simple shear in two dimensions. Inserting the expansion \eqref{eq:B-expansion-def} into \eqref{eq:g}, we obtain at steady state
\begin{equation}
     \frac{1}{2}Pe \frac{\partial g_n}{\partial \phi}  + \frac{\partial^2 g_n}{\partial \phi^2} = \frac{1}{2}Pe \frac{\partial }{\partial \phi}\left[ \cos(2\phi) g_{k-1}\right],
\end{equation}
where $k=1,2,\cdots$. The normalization becomes $\avg{g_0} =1$ and $\avg{g_k}=0$ for $k=1,2,\cdots$.

The first four terms in the expansion are written as 
\begin{subequations}
    \begin{align}
        &g_0(\phi) =1,\quad g_1(\phi)=a_{11}\cos(2\phi) + a_{12}\sin(2\phi),\\ 
        & g_2(\phi)=a_{21}\cos (4 \phi )+a_{22}\sin (4 \phi ),\\ 
        & g_3(\phi) = a_{31} \cos(2\phi) + a_{32}\sin(2\phi) + a_{33}\cos(6\phi) + a_{34}\sin(6\phi),
    \end{align}
\end{subequations}
where 
\begin{subequations}
    \begin{align}
        a_{11}&= \frac{Pe^2}{Pe^2+16},  \quad a_{12}= \frac{4 Pe }{Pe^2+16},\\ 
        a_{21}&=\frac{Pe^2 \left(Pe^2-32\right) }{2 \left(Pe^2+16\right) \left(Pe^2+64\right)},\quad a_{22}=\frac{6 Pe^3 }{\left(Pe^2+16\right) \left(Pe^2+64\right)}, \\ 
       a_{31}&= \frac{Pe^4 \left(Pe^2-80\right)}{4 \left(Pe^2+16\right)^2 \left(Pe^2+64\right)},\\
       a_{32}&=-\frac{Pe^3 \left(-16 Pe^4-2176 Pe^2+18432\right) }{4 \left(Pe^2+16\right)^2 \left(Pe^2+64\right) \left(Pe^2+144\right)}, \\ 
       a_{33}&= -\frac{Pe^3 \left(2816 Pe-Pe^3 \left(Pe^2-160\right)\right) }{4 \left(Pe^2+16\right)^2 \left(Pe^2+64\right) \left(Pe^2+144\right)}, \\ 
       a_{34}&= -\frac{Pe^3 \left(6144-24 Pe^4\right) }{4 \left(Pe^2+16\right)^2 \left(Pe^2+64\right) \left(Pe^2+144\right)}. 
    \end{align}
\end{subequations}

The solution at $O(B^4)$ is written as 
\begin{equation}
    g_4(\phi) = a_{41} \cos(4\phi) + a_{42}\sin(4\phi) + a_{43}\cos(8\phi) + a_{44}\sin(8\phi), 
\end{equation}
where
\begin{subequations}
    \begin{align}
        a_{41} &= \frac{Pe^2 (a_{31}+a_{33})-8Pe (a_{32}+a_{34})}{2 \left(Pe^2+64\right)}, \\ 
        a_{42}&= \frac{8Pe (a_{31}+a_{33})+Pe^2 (a_{32}+a_{34})}{2 \left(Pe^2+64\right)}, \\ 
        a_{43}&= \frac{Pe (Pe\,a_{33} -16 a_{34})}{2 \left(Pe^2+256\right)}, \quad a_{44}= \frac{Pe (16 a_{33}+Pe\, a_{34} )}{2 \left(Pe^2+256\right)}. 
    \end{align}
\end{subequations}
The solution at $O(B^5)$ is given by 
\begin{eqnarray}
    g_5(\phi) = && a_{51} \cos (2 \phi )+a_{52} \sin (2 \phi )+a_{53} \cos (6 \phi )\nonumber\\
    && +a_{54} \sin (6 \phi )+a_{55} \cos (10 \phi )+a_{56} \sin (10 \phi )
\end{eqnarray}
where 
\begin{subequations}
    \begin{align}
        a_{51} &= \frac{Pe (a_{41} Pe-4 a_{42})}{2 \left(Pe^2+16\right)}, \quad a_{52}=\frac{Pe (4 a_{41}+a_{42} Pe)}{2 \left(Pe^2+16\right)},\\ 
        a_{53}&= \frac{Pe^2 (a_{41}+a_{43})-12Pe (a_{42}+a_{44})}{2 \left(Pe^2+144\right)}, \quad a_{54}=\frac{12Pe (a_{41}+a_{43})+Pe^2 (a_{42}+a_{44})}{2 \left(Pe^2+144\right)},\\ 
        a_{55}&=\frac{Pe (a_{43} Pe-20 a_{44})}{2 \left(Pe^2+400\right)}, \quad a_{56}=\frac{Pe (20 a_{43}+a_{44} Pe)}{2 \left(Pe^2+400\right)}. 
    \end{align}
\end{subequations}
The solution at $O(B^6)$ is given by 
\begin{eqnarray}
    g_6(\phi) = && a_{61} \cos (4 \phi )+a_{62} \sin (4 \phi )+a_{63} \cos (8 \phi )\nonumber\\
    && +a_{64} \sin (8 \phi )+a_{65} \cos (12 \phi )+a_{66} \sin (12 \phi )
\end{eqnarray}
where 
\begin{subequations}
    \begin{align}
        a_{61}&=\frac{Pe^2 (a_{51}+a_{53})-8Pe (a_{52}+a_{54})}{2 \left(Pe^2+64\right)}, \quad a_{62}=\frac{8Pe (a_{51}+a_{53})+Pe^2 (a_{52}+a_{54})}{2 \left(Pe^2+64\right)},\\ 
        a_{63}&=\frac{Pe^2 (a_{53}+a_{55})-16Pe (a_{54}+a_{56})}{2 \left(Pe^2+256\right)}, \quad a_{64}=\frac{16Pe (a_{53}+a_{55})+Pe^2 (a_{54}+a_{56})}{2 \left(Pe^2+256\right)},\\ 
        a_{65}&=\frac{Pe (a_{55} Pe-24 a_{56})}{2 \left(Pe^2+576\right)}, \quad a_{66}=\frac{Pe (24 a_{55}+a_{56} Pe)}{2 \left(Pe^2+576\right)}. 
    \end{align}
\end{subequations}

Similarly, the displacement field can be solved order by order. At $O(B^n)$, equation \eqref{eq:beq} is given by 
\begin{equation}
    \label{eq:beq-expansion}
    \frac{\partial }{\partial \phi}\left( \Omega_0 b_n + \Omega_1 b_{n-1} - \frac{\partial b_n}{\partial \phi}\right) = 2 \frac{\partial g_n}{\partial \phi} + \sum_{i= n-j}\sum_{j} \Omega_i^\mathrm{eff} g_j - \left(  \Omega_0 g_n + \Omega_1 g_{n-1}\right),
\end{equation}
where $\Omega_0 = -\frac{1}{2}Pe$,  $\Omega_1 = \frac{1}{2}Pe\cos(2\phi)$ and $n=0,1,\cdots$. In \eqref{eq:beq-expansion}, $b_k=0$ for $k<0$. At $O(1)$, we have 
\begin{equation}
    \label{eq:b0-eq}
    \frac{\partial }{\partial \phi}\left( \Omega_0 b_0  - \frac{\partial b_0}{\partial \phi}\right) = 2 \frac{\partial g_0}{\partial \phi} + \left( \Omega_0^\mathrm{eff}-\Omega_0\right)g_0.
\end{equation}
Since $g_0 =1$, we have $b_0=0$. At $O(B)$, we have 
\begin{equation}
   \frac{\partial }{\partial \phi}\left( \Omega_0 b_1 +\Omega_1 b_0  - \frac{\partial b_1}{\partial \phi}\right) = 2 \frac{\partial g_1}{\partial \phi} + \left( \Omega_0^\mathrm{eff}-\Omega_0\right)g_1+ \left( \Omega_1^\mathrm{eff}-\Omega_1\right)g_0.
\end{equation}
The solution is given by 
\begin{equation}
    b_1(\phi)=\frac{Pe}{2 \left(Pe^2+16\right)^2}\left[ Pe \left(Pe^2-48\right) \sin (2 \phi )+\left(64-12 Pe^2\right) \cos (2 \phi )  \right].
\end{equation}
At $O(B^2)$, we have 
\begin{align}
    b_2(\phi) = \frac{Pe^2}{8 \left(Pe^2+16\right)^2 \left(Pe^2+64\right)^2}\Big[ -4 Pe \left(17 Pe^4+208 Pe^2-19456\right) \cos (4 \phi )\nonumber \\ 
     + 3 \left(Pe^6-144 Pe^4-6656 Pe^2+32768\right) \sin (4 \phi ) \Big].
\end{align}

Following this procedure, we have 
\begin{subequations}
\begin{align}
       b_3(\phi) &= c_{31} \cos(2\phi) + c_{32}\sin(2\phi) + c_{33}\cos(6\phi) + c_{34}\sin(6\phi),  \\ 
       b_4(\phi) &= c_{41} \cos(4\phi) + c_{42}\sin(4\phi) + c_{43}\cos(8\phi) + c_{44}\sin(8\phi),\\
       b_5(\phi) &= c_{51} \cos (2 \phi )+c_{52} \sin (2 \phi )+c_{53} \cos (6 \phi )\nonumber\\
     & \quad +c_{54} \sin (6 \phi )+c_{55} \cos (10 \phi )+c_{56} \sin (10 \phi ),\\ 
     b_6(\phi)&= c_{61} \cos (4 \phi )+c_{62} \sin (4 \phi )+c_{63} \cos (8 \phi )\nonumber\\
    &\quad  +c_{64} \sin (8 \phi )+c_{65} \cos (12 \phi )+c_{66} \sin (12 \phi ), 
\end{align}
\end{subequations}
where $c_{ij}$ can be readily obtained by inserting the expressions into \eqref{eq:beq-expansion}.

\section{Asymptotic analysis for $1-\alpha = B(1+\alpha)$ in the strong flow limit}
\label{sec:appendix-generic-boundary}
We develop an asymptotic solution in the large-$Pe$ limit for a general linear flow under the constraint that $1-\alpha = B(1+\alpha)$. Under this condition, we have $\Omega\tau_R = -Pe(1-\alpha) \sin^2\phi$. For convenience, we define $\epsilon = 1/(Pe(1-\alpha))$ and consider the behaviour of the macrotransport equations when $\epsilon \ll 1$. We note that $\alpha \neq 1$ is assumed.

The dimensionless average field at steady state, $g=g(\phi)$,  is governed by 
\begin{equation}
    \label{eq:g-boundary}
    \frac{\odiff }{\odiff\phi}( \sin^2\phi g) +\epsilon \frac{\odiff^2 g}{\odiff\phi^2}=0.
\end{equation}
Noting that $g(\phi+\pi) = g(\phi)$, we consider $\phi \in [-\pi/2, \pi/2]$. The normalization is given by 
\begin{equation}
    \label{eq:g-boundary-norm}
    \frac{1}{\pi}\int_{-\pi/2}^{\pi/2} g(\phi)\odiff\phi =1. 
\end{equation}
An asymptotic solution to equation \eqref{eq:g-boundary}  subject to the normalization condition \eqref{eq:g-boundary-norm} has been developed by \citet{brenner1970orientation}. Since the solution to $g$ is required for the consideration of the displacement field $b$, we first repeat the calculations of \citet{brenner1970orientation} for $g$ and then proceed to consider $b$. 

\begin{figure}
  \centerline{\includegraphics[width=5in]{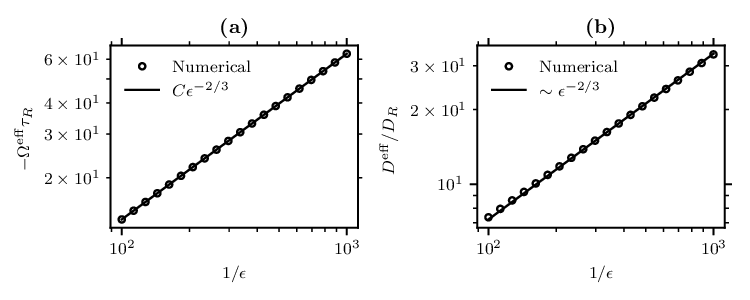}}
  \caption{(a) Plots of the numerical and asymptotic solutions to the average angular drift velocity as a function of $1/\epsilon$ for $B=1,\alpha=0$. (b) Plots of the numerical solutions to the non-dimensional effective long-time dispersion coefficient as a function of $1/\epsilon$ for $B=1,\alpha=0$. The solid line is given by $D^\mathrm{eff}/D_R = \tilde{C} \epsilon^{-2/3}$, where $\tilde{C}$ is obtained from numerical data. }
\label{fig:appB}
\end{figure}

Integrating \eqref{eq:g-boundary} once, we obtain 
\begin{equation}
\label{eq:g-boundary-1st-order}
    \sin^2\phi g +\epsilon \frac{\odiff g}{\odiff\phi}=C^\prime(\epsilon), 
\end{equation}
where we remark that $C^\prime$ only depends on $\epsilon$. In the bulk, we have $\sin^2\phi \;g \sim C^\prime$ or $g\sim C^\prime/\sin^2\phi$. This bulk result breaks down when $\sin^2\phi\, g \sim \epsilon \, \odiff g/(\odiff\phi)$ or $\phi =O(\epsilon^{1/3})$, which suggests the existence of a boundary layer with thickness $O(\epsilon^{1/3})$ at $\phi=0$. In the boundary layer, from \eqref{eq:g-boundary-1st-order}, we obtain $g\sim C^\prime \epsilon^{-2/3} $. Using the normalization condition \eqref{eq:g-boundary-norm}, one can show that $C^\prime = O(\epsilon^{1/3})$. Following \citet{brenner1970orientation}, we define $C^\prime = C \epsilon^{1/3}$, where $C=O(1)$. In the bulk, we define $g(\phi) = C\epsilon^{1/3} H(\phi; \epsilon)$. Introducing the stretched coordinate $s = \phi/\epsilon^{1/3}$, we define $g(s) = C\epsilon^{-1/3} Q(s;\epsilon)$ in the boundary layer. Expanding $H$ and $Q$, we write 
\begin{equation}
    H = H_0 +\epsilon H_1+\cdots, \quad Q = Q_0 + \epsilon^{2/3}Q_1+\cdots, 
\end{equation}
where $H_0 = 1/\sin^2\phi$ and $Q_0$ is governed by 
\begin{equation}
    s^2 Q_0(s) + \frac{\odiff Q_0}{\odiff s} =1.
\end{equation}
The solution to $Q_0$ is given by 
\begin{equation}
    Q_0(s) = e^{-s^3/3}\int_{-\infty}^s e^{r^3/3}\odiff r.
\end{equation}
To leading order, the normalization condition reduces to 
\begin{equation}
    C \int_{-\infty}^{+\infty} Q_0(s) \odiff  s =\pi, 
\end{equation}
which allows us to obtain 
\begin{equation}
    C = \frac{3^{5/6} \pi }{6 \sqrt{3} \left(\Gamma \left(4/3\right)\right)^2+2^{4/3} \sqrt{\pi } \Gamma \left(7/6\right)} = \frac{3^{1/3}\pi}{[\Gamma(1/3)]^2},
\end{equation}
where $\Gamma(z) = \int_0^\infty t^{z-1}e^{-t}\odiff t$ is the gamma function and $C\approx 0.6313$. We can construct a leading-order composite solution to $g$ as $ C \epsilon^{-1/3} Q_0(\phi/\epsilon^{1/3})$. From this, we obtain the leading-order drift as $\Omega^\mathrm{eff}\tau_R= -C \epsilon^{-2/3} +o(\epsilon^{-2/3})$. In figure \ref{fig:appB}(a) we compare the asymptotic result with the results from numerical solutions of the macrotransport equations for $B=1$ and $\alpha=0$ in the large-$Pe$ regime. 

The displacement field is governed by 
\begin{equation}
    - \frac{\odiff }{\odiff\phi}\left(\sin^2\phi\, b  \right) - \epsilon \frac{\odiff^2 b}{\odiff\phi^2} = 2 \epsilon \frac{\odiff g }{\odiff\phi} + \epsilon \left(\Omega^{*\mathrm{eff}} - \Omega^*\right)g, 
\end{equation}
where $\Omega^{*\mathrm{eff}} = \Omega^\mathrm{eff}\tau_R$ and $\Omega^* = \Omega\tau_R$. Noting that the dominant term on the right-hand side outside the boundary layer is $\epsilon\Omega^*g=O(\epsilon^{1/3})$, we have $b = O(\epsilon^{1/3})$ in the bulk. In the boundary layer, a dominant balance reveals that $b=O(\epsilon^{-1/3})$. While the resulting equations are not analytically tractable, the above analysis allows us to show that $D^\mathrm{eff}/D_R = O(\epsilon^{-2/3})$ as $\epsilon \to 0$.  In figure \ref{fig:appB} we present the dispersion coefficient obtained from numerical solutions of the macrotransport equations for $B=1$ and $\alpha=0$ in the large-$Pe$ regime. The solid line has the form  $D^\mathrm{eff}/D_R = \tilde{C} \epsilon^{-2/3}$, where $\tilde{C}$ is a numerical constant that fits the data closely.

\bibliographystyle{jfm}
\bibliography{refs}

\end{document}